\documentclass[10pt,conference]{IEEEtran}
\usepackage{amsfonts}
\usepackage{setspace}

\usepackage{setspace}
\usepackage{amsmath}
\usepackage{amssymb}
\usepackage[dvips]{graphicx}
\usepackage{epsfig}
\usepackage[ps2pdf,
bookmarks=false,
bookmarksnumbered=false, 
bookmarksopen=false, 
colorlinks=false]{}

\newcommand{\tsnr}{{\text{\footnotesize{SNR}}}}

\newcommand{\E}{\mathbb{E}}

\newcommand{\bDT}{{\mathbf{DT}}}
\newcommand{\bDR}{{\mathbf{DR}}}

\newcommand{\BS}{{\mathbf{BS}}}
\newcommand{\CU}{{\mathbf{CU}}}

\newtheorem{Rem}{Remark}

\begin{document}

%
\title{Joint Mode Selection and Resource Allocation for D2D Communications via Vertex Coloring}

\author{Yi Li, M. Cenk Gursoy, Senem Velipasalar, Jian Tang
\\Department of Electrical Engineering and Computer Science,
Syracuse University, Syracuse, NY 13244
\\Email: yli33@syr.edu, mcgursoy@syr.edu, svelipas@syr.edu, jtang02@syr.edu}

\maketitle

\begin{abstract}
Device-to-device (D2D) communication underlaid with cellular networks is a new paradigm, proposed to enhance the performance of cellular networks. By allowing a pair of D2D users to communicate directly and share the same spectral resources with the cellular users, D2D communication can achieve higher spectral efficiency, improve the energy efficiency, and lower the traffic delay. In this paper, we propose a novel joint mode selection and channel resource allocation algorithm via the vertex coloring approach. We decompose the problem into three subproblems and design algorithms for each of them. In the first step, we divide the users into groups using a vertex coloring algorithm. In the second step, we solve the power optimization problem using the interior-point method for each group and conduct mode selection between the cellular mode and D2D mode for D2D users, and we assign channel resources to these groups in the final step. Numerical results show that our algorithm achieves higher sum rate and serves more users with relatively small time consumption compared with other algorithms. Also, the influence of system parameters and the tradeoff between sum rate and the number of served users are studied through simulation results.
\end{abstract}

\thispagestyle{empty}

\section{Introduction}
Device-to-Device (D2D) communication underlaid with cellular networks is a new paradigm for next-generation 5G wireless systems. D2D communication enables users to communicate directly without going through the base station, and potentially reuse the same spectral resources with cellular users. In a cellular network, D2D users can transmit directly using a dedicated frequency band or by sharing the spectrum with cellular users, and they can also transmit in the same way as cellular users via the base station. The advantages of D2D communications were studied in \cite{D2D_KB}, and it was shown that D2D communication could greatly enhance the spectral efficiency and lower the latency. A comprehensive overview was provided in \cite{D2D_survey}, where different modeling assumptions and key considerations in D2D communications were detailed.

Mode selection and resource allocation are two key problems in D2D communication, which has attracted much interest. In mode selection, each D2D user has to decide whether to communicate directly in the D2D mode, or transmit via a D2D two-hop channel through the base station in the cellular mode. In resource allocation, the system has to assign a channel resource to each user, and users have to optimize their transmission power. The resource allocation problem in D2D cellular networks is rather complicated because D2D users can reuse (i.e., share) the same channel resources with cellular users and inflict interference to them. Due to this reusing mechanism, the number of possible solutions for channel assignment increases exponentially with the number of D2D users, and the power optimization problem becomes high-dimensional and non-convex. Therefore, the analysis becomes even more complicated when mode selection and resource allocation problems are considered jointly for improved performance.

In the literature, many studies have been conducted to address the mode selection and resource allocation problems for D2D cellular networks. For instance, the authors of \cite{mode_selection_DK} considered the mode selection problem in a cell with one D2D pair and one cellular user. In \cite{matching_F}, a channel assignment algorithm for an uplink reuse mode was proposed via bipartite matching approach, and it was further extended for both uplink and downlink reuse in \cite{matching_JH}. More recently, the joint mode selection and resource allocation in a general cellular network with multiple D2D pairs were addressed in \cite{Joint_selection}. In order to reduce the complexity in analysis, most of these studies were based on the instantaneous channel conditions. In such cases, the system may have to perform the mode selection and resource allocation very frequently, resulting in high computational load and significant cost. Unlike these works, we have recently analyzed the performance based on average throughput rather than instantaneous rate/capacity values. We first studied the mode selection and resource allocation for a simple model with one user and one D2D pair in \cite{li2016device}, and then we solved the joint mode selection and resource allocation problem for a more general network model with multiple cellular and D2D users in \cite{li2016matching} using a matching algorithm. However, our results still rely on the assumption that each channel cannot be reused by more than two transmission links. Without this assumption, our previous algorithms become overly time consuming.

In order to achieve improved results with lower time consumption, several algorithms were proposed via game-theoretic approaches. For example, the resource allocation problem was considered in \cite{D2D_ICA} via the reverse iterative combinatorial auction game, and the authors of \cite{li2014coalitional} solved a similar problem using the coalitional game theory. Besides the game-theoretic techniques, vertex coloring is another method that can efficiently divide D2D users into groups in which interference constraints are satisfied. In \cite{vertex_lte}, vertex coloring algorithm was used to group D2D users with the goal of avoiding interference. A similar approach was used in \cite{vertex_lte} and \cite{vertex_lee} to maximize the instantaneous sum rate while satisfying the instantaneous SINR constraints, and a frequency band assignment process was also included after dividing D2D users into groups in \cite{vertex_efficient}.

The main contributions of this work are given as follows:
\begin{enumerate}
  \item In our analysis, we divide the problem into three subproblems, namely user partition, power allocation and channel assignment. Different from prior works, the power allocation, mode selection and channel assignment are considered after grouping D2D users via vertex coloring.
  \item Algorithms are designed for each subproblem, and we propose a novel three-step joint mode selection and resource allocation method by combining these algorithms designed for the three subproblems.
  \item We incorporate the adaptation of the interference constraints in the grouping step when the given interference constraints are relatively loose.
  \item Fairness among the users in the same group is also considered in the power allocation step.
  \item Further comparisons are made via simulations, and the influence of system parameters is investigated via numerical results.
\end{enumerate}

\section{System Model and Assumptions}
\begin{figure}
\begin{center}
\includegraphics[width=0.4\textwidth]{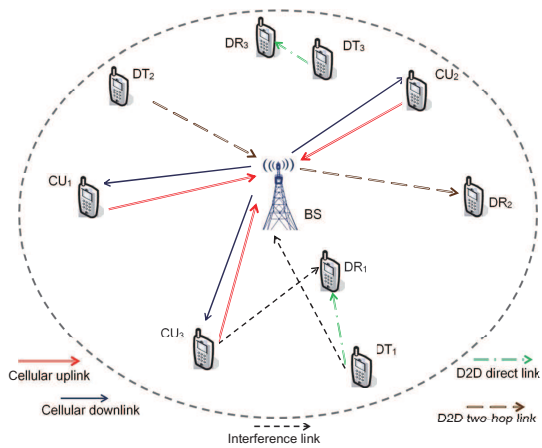}
\caption{System model}\label{fig:system}
\vspace{-0.5cm}
\end{center}
\end{figure}

In this work, as shown in Figure \ref{fig:system}, we consider a D2D underlaid cellular network, which has one base station ($\BS$), $N_c$ cellular users \{$\CU_1,\CU_2,\cdots,\CU_{N_c}$\} and $N_d$ D2D pairs $\{(\bDT_1,\bDR_1),(\bDT_2,\bDR_2),\cdots,(\bDT_{N_d},\bDR_{N_d})\}$. We assume that the D2D transmission is one-way, in which $\bDT_i$ and $\bDR_i$ represent the transmitter and receiver of the $i^{\text{th}}$ D2D pair, respectively. Each D2D pair can choose between the cellular mode and D2D mode. In D2D mode, D2D users transmit through D2D direct links, while in cellular mode, they transmit via D2D two-hop links through the base station. Each cellular user transmits to the base station through an uplink channel, and receives data from the base station via a downlink channel. Hence, there are overall $N_c$ uplinks, $N_c$ downlinks and $N_d$ D2D links. The maximum transmission power of a cellular user and D2D transmitter are set at $P_c$ and $P_d$, respectively. When acting as a transmitter, the maximum transmission power of the base station is $P_b$ in each channel. Therefore, the overall transmission power of the base station depends on the number of cellular users and the number of D2D pairs operating in the cellular mode.

There are $N$ available orthogonal channels for this cellular network, each of them having a bandwidth of $B$. For simplicity, there are four assumptions regarding the channel allocation, which were also made in many related works such as \cite{matching_F} and \cite{li2016matching}:
\begin{enumerate}
  \item A D2D pair operating in the cellular mode cannot share its channel with other users.
  \item Each cellular link, including both uplink and downlink, is allocated a single orthogonal channel, and channels cannot be shared by different cellular links.
  \item It is necessary for a pair of direct links to satisfy the pair-wise interference constraints given below in (\ref{eq:pair_interf}) in Section \ref{sec:partition} to reuse the same channel.
  \item Each link, including D2D direct link, D2D two-hop link, cellular uplink and downlink, can operate in one channel at most.
  \item The base station has the knowledge of the distributions of all channel fading coefficients, i.e., has statistical channel side information.
\end{enumerate}
The first assumption helps to protect the performance of those D2D users that select the cellular mode. In general, D2D users that select the cellular mode usually have weak connections to their corresponding receivers, i.e., the distances between D2D transmitter, D2D receiver and the base station are relatively large. Therefore, assigning these D2D two-hop channels dedicated transmission resources provides a certain level of quality of service (QoS) guarantee. The second assumption guarantees the performance of cellular users, which have higher priorities than D2D users. The third assumption controls the interference among the users that reuse the same transmission resource. The last assumption implies that our resource allocation algorithm is performed at the base station, and our algorithm only requires the knowledge of the fading distributions, i.e., statistical channel side information. In general, fading distributions depend on the environment and distance between the transmitter and receiver. If a certain fading model is considered, such as Rayleigh, Rician or Nakagami-$m$ fading, then the fading distributions are mainly determined by the location of the users.

According to these assumptions, a channel can be assigned to a single D2D link, a single cellular link, a group of D2D direct links or a group of D2D direct links together with a cellular link. For the last two cases, the users transmitting in the same channel cause interference to each other. Note that in some systems, cellular downlinks do not share transmission resources with D2D users. In such cases, we just need to first assign transmission resources to those cellular downlinks before applying our algorithm. In this paper, we assume that $2N_c\leq N \leq 2N_c+N_d$, which implies that we have sufficient number of channels to guarantee the performance requirements of all cellular users. However, having dedicated channels for all D2D users is not feasible in such a situation, and reusing/sharing of channel resources has to be considered in order to serve as many D2D users as possible.

The channels are assumed to experience ergodic fading, and the fading coefficients are denoted by $h$.
Fading coefficients in different frequency bands are assumed to be independent and identically distributed (i.i.d.). In the following analysis, the magnitude-squares of the fading coefficients are denoted by $z=|h|^2$. At each receiver, the background noise is assumed to follow an independent complex Gaussian distribution with zero mean and variance $\sigma^2$, i.e., $n \sim\mathcal{CN}(0,\sigma^2)$. Therefore, the SNR of each transmitter can be defined as $\tsnr=\frac{P}{B\sigma^2}$, where $P$ represents the transmission power.

In this work, we consider mode selection, power optimization and channel allocation jointly to maximize the throughput as well as the number of users served in the network. In the next section, we introduce our algorithm step by step.

\section{Joint Mode Selection and Resource Allocation Algorithm}
In this section, we introduce our three-step joint mode selection and resource allocation algorithm in detail. In the first step, we divide the transmission links into groups via the vertex coloring method. In the second step, we conduct power optimization for each group, and perform mode selection between D2D mode and cellular mode for those D2D links which form groups. In the last step, we assign channels to those groups.

Before applying the algorithm, we enumerate cellular uplinks from $1$ to $N_c$, cellular downlinks from $N_c+1$ to $2N_c$, and D2D direct links from $2N_c+1$ to $2N_c+N_d$. D2D two-hop links are only considered in the mode selection part in the second step. With the given link indices, we can denote the magnitude-square of the fading coefficient between the transmitter of link $i$ and the receiver of link $j$ by $z_{i,j}$, and we can represent the \emph{expected} values of $z$ collectively in a channel fading matrix $\bold{Z}$.

Two main objectives of our algorithm are to maximize the sum rate and to maximize the number of users served in the network. Most of the time, these two goals cannot be achieved simultaneously because of the presence of interference. In the following discussion, we illustrate how to balance these two goals via parameter selection.
\subsection{Partition via Vertex Coloring Method}\label{sec:partition}
The first step of our algorithm is transmission link partition. The partition algorithm divides transmission links into small groups, greatly reducing the dimensionality of the power optimization problem in the second step. According to our channel assignment assumptions, multiple cellular links cannot be in the same group, and any two links in the same group have to satisfy the pair-wise interference constraints given by
\begin{align}
\begin{cases}
P_{i max}\bar{z}_{ii}/(P_{j max}\bar{z}_{ji})\geq \gamma \\
P_{j max}\bar{z}_{jj}/(P_{i max}\bar{z}_{ij})\geq \gamma
\end{cases}\label{eq:pair_interf}
\end{align}
where $P_{i max}$ and $P_{j max}$ are the maximum transmission powers over links $i$ and $j$ respectively, $\bar{z}$ represents the expected value of $z$, and $\gamma$ is the interference threshold. These pair-wise interference constraints provide QoS guarantees for both cellular and D2D users from the perspective of interference control.

The key steps of our partition algorithm are to construct a graph while regarding these $2N_c+N_d$ transmission links as vertices, and to perform the partition using the minimum vertex coloring algorithms from graph theory. Note that these algorithms divide all vertices into minimum number of groups such that any two vertices in the same group are not connected. Therefore, we construct the graph by checking each pair of vertices, and connect them if they cannot be in the same group. A detailed description of our partition algorithm is given in Table \ref{Alg:Partition}. The output of this algorithm is a partition with size $n_g$, and each element of the partition is a set of vertices that form a group.
\begin{table}
\footnotesize
\caption{\label{Alg:Partition}Algorithm 1}
\begin{tabular}{p{8.4cm}}
\hline
\hline
Partition Algorithm\\
\hline
\hline
\textbf{Input:} interference threshold $\gamma$, channel fading matrix $\bold{Z}$.\\
\hline
\textbf{Output:} partition $\bold{\Pi}={\pi_1,\pi_2,\cdots,\pi_{n_g}}$.\\
\hline
\textbf{For} $i=1:2N_c+Nd$\\
\hspace{0.5cm} $\gamma_i=\gamma$;\\
\textbf{End}\\
Generate a random permutation of integers from $1$ to $2N_c+N_d$, and denote it by $\mathbb{A}_1$;\\
\textbf{For each} $i\in \mathbb{A}_1$\\
\hspace{0.5cm}Generate a random permutation of integers from $i+1$\\ \hspace{0.5cm}to $2N_c+N_d$, and denote it by $\mathbb{A}_2$;\\
\hspace{0.5cm}\textbf{For each} $j\in\mathbb{A}_2$\\
\hspace{1cm}\textbf{If} both links $i$ and $j$ are smaller than $2N_c$\\
\hspace{1.5cm}Create an edge between vertices $i$ and $j$;\\
\hspace{1cm}\textbf{Elseif} links $i$ and $j$ cannot satisfy
\begin{align*}
\begin{cases}
P_{i max}\bar{z}_{ii}/(P_{j max}\bar{z}_{ji})\geq \gamma_i \\
P_{j max}\bar{z}_{jj}/(P_{i max}\bar{z}_{ij})\geq \gamma_j
\end{cases}
\end{align*}
\hspace{1.5cm}Create an edge between vertices $i$ and $j$;\\
\hspace{1cm}\textbf{Else}\\
\hspace{1.5cm} Increase both $\gamma_i$ and $\gamma_j$ by $\Delta\gamma$;\\
\hspace{1cm}\textbf{End}\\
\hspace{0.5cm}\textbf{End}\\
\textbf{End}\\
Apply the Welsh-Powell algorithm to get the partition $\bold{\Pi}$;\\
\hline
\hline
\end{tabular}
\end{table}
In order to further control the interference and number of users in a group, we gradually increase the $\gamma$ values of each link. As we can see in the algorithm, all threshold values are set at $\gamma$ initially. Each time we find a pair of links that can be in the same group, we increase the thresholds of these two links by $\Delta\gamma$. This mechanism can effectively limit the received interference at each receiver, and balance the size of each group. Also due to this mechanism, two links may have a higher chance to be in the same group if we check them earlier. In order to let the vertices to have equal chances to connect with each other, we use random orders to choose link pairs in the double for-loop. In the last step of the algorithm, we use the Welsh-Powell algorithm \cite{welsh1967upper} to solve the vertex coloring problem. Welsh-Powell algorithm is a very fast algorithm that can provide good results effectively.

In this step, $\gamma$ and $\Delta\gamma$ are the parameters to control the tradeoff between sum rate and number of users served by the system. For large values of $\gamma$ and $\Delta\gamma$, the interference is well controlled, but the system serves potentially small number of users. On the other hand, for small $\gamma$ and $\Delta\gamma$ values, more users can reuse the same channel resource, but the interference may lower the sum rate.

In practice, the partition algorithm can potentially provide us a partition with size smaller than the number of channels, which means that some of the channels will not be utilized, because each user group $\pi_i$ is assigned a channel in the third step of our algorithm. In order to avoid this situation, we need to further improve our partition algorithm using a $\gamma$-adjusting algorithm described in Table \ref{Alg:gama_adj}. In Algorithm 2, we find a threshold $\hat{\gamma}$ that makes the partition size $n_g=N$ through bisection search. Notice that the threshold value that can achieve $n_g=N$ is not unique, and the time consumption of this adjusting algorithm is very small.
\begin{table}
\footnotesize
\caption{\label{Alg:gama_adj}Algorithm 2}
\begin{tabular}{p{8.4cm}}
\hline
\hline
$\gamma$ Adjusting Algorithm\\
\hline
\hline
\textbf{Input:} interference threshold $\gamma$, channel fading matrix $\bold{Z}$.\\
\hline
\textbf{Output:} partition $\bold{\Pi}={\pi_1,\pi_2,\cdots,\pi_{n_g}}$.\\
\hline
Run Algorithm 1 with threshold $\gamma$;\\
\textbf{If} $n_g\geq N$\\
\hspace{0.5cm}End process;\\
\textbf{End}\\
Set $\hat{\gamma}=\gamma$;\\
\textbf{While} $n_g<N$\\
\hspace{0.5cm}$\hat{\gamma}=2\hat{\gamma}$;\\
\hspace{0.5cm}Run Algorithm 1 with threshold $\hat{\gamma}$;\\
\textbf{End}\\
Set the upper bound $\gamma_u=\hat{\gamma}$, lower bound $\gamma_l=\hat{\gamma}/2$, and $\hat{\gamma}=(\gamma_u+\gamma_l)/2$;\\
\textbf{While} $n_g\neq N$\\
\hspace{0.5cm}Run Algorithm 1 with threshold $\hat{\gamma}$;\\
\hspace{0.5cm}\textbf{If} $n_g>N$\\
\hspace{1cm} $\gamma_u=\hat{\gamma}$;\\
\hspace{0.5cm}\textbf{Elseif} $n_g<N$\\
\hspace{1cm} $\gamma_l=\hat{\gamma}$;\\
\hspace{0.5cm}\textbf{End}\\
\hspace{0.5cm}$\hat{\gamma}=(\gamma_u+\gamma_l)/2$;\\
\textbf{End}\\
\hline
\hline
\end{tabular}
\end{table}

After obtaining the partition, we conduct power optimization and mode selection in the second step.
\subsection{Power Optimization and Mode Selection}
In the second step, we do power optimization for each group. If a group just contains a single D2D direct link, then we perform mode selection for this D2D pair.

~\\
\textbf{Power Optimization}

If a group only contains one direct link, then the transmitter transmits with its maximum power. For the groups containing multiple transmission links, a general expression of the objective function for the power optimization problem in group $\pi_i$ is
\begin{equation}\label{eq:obj}
Obj(\bold{P_i})=\sum_{k} \omega_k CRT_k(\bold{P_i}),
\end{equation}
where $\bold{P_i}$ represents the power vector which consists of the transmission powers of the transmitters in group $\pi_i$, the function $CRT$ can be defined based on the criteria selected in the optimization problem, such as the maximization of the sum rate, energy efficiency, or minimum rate, and $\omega_k$ is the corresponding weight of $CRT_k$ which indicates the significance of $CRT_k$. The formulation given in (\ref{eq:obj}) can provide QoS and fairness guarantees. For instance, by choosing the energy efficiency as a criterion, a certain energy efficiency performance can be achieved; or by choosing the minimum rate as a criterion, the minimum rate performance of each users can be guaranteed.

In this work, we consider both the sum rate and minimum rate as the criteria, and formulate our power optimization problem for group $\pi_i$ as
\begin{align}
&\textbf{Maximize}_{\;\bold{P_i}}\hspace{0.3cm} \sum_{j\in \pi_i} \E\{R_j(\bold{P_i})\}+\mu\min_{j\in\pi_i}\big\{\E\{R_j(\bold{P_i})\}\big\}\\
&\textbf{Subject to}\hspace{1cm}0\leq P_j\leq P_{jmax},\; \text{for}\; j\in\pi_i
\end{align}
where $P_{jmax}$ represents the maximum transmission power in link $j$. The evaluation of the average transmission rate $\E\{R_j(\bold{P_i})\}$ is discussed in Remark \ref{Rem1} below. In this problem, $\mu$ is the weight parameter for the minimum rate. For small $\mu$ values, the objective function is mainly determined by the sum rate component, which may sacrifice the rates of some users. On the other hand, for large $\mu$ values, the objective function is mainly dominated by the minimum rate component, which may limit the sum rate. This optimization problem can be transformed into
\begin{align}
&\textbf{Maximize}_{\;\bold{P_i},r}\hspace{0.3cm} \sum_{j\in \pi_i} \E\{R_j(\bold{P_i})\}+\mu r\\
&\textbf{Subject to}\hspace{1cm}0\leq P_j\leq P_{jmax},\; \text{for}\; j\in\pi_i\\
&\hspace{2.5cm}\E\{R_j(\bold{P_i})\}\geq r, \; \text{for}\; j\in\pi_i
\end{align}
for which suboptimal solutions can be obtained via the interior-point method \cite{boyd2004convex}. In order to improve the performance, we need to repeat the algorithm several times with randomly selected initial points.
\begin{Rem}\label{Rem1}
In order to determine the average rate of a user accurately and efficiently,we perform numerical integration. To evaluate this high-dimensional integral, we transform it into two single integrals for certain specific fading models:
\begin{align}
\E\{R_j(\bold{P_i})\}=&\E\left\{B\log_2\left(1+\frac{\tsnr_j z_{jj}}{1+\sum_{k\in\pi_i, k\neq j}\tsnr_k z_{kj}}\right)\right\}\\
                     =&\E\left\{B\log_2\left(1+\sum_{k\in\pi_i}\tsnr_k z_{kj}\right)\right\}\notag\\
                     &\hspace{0.5cm}-\E\left\{B\log_2\left(1+\sum_{k\in\pi_i, k\neq j}\tsnr_k z_{kj}\right)\right\}.\label{eq:int_simplify}
\end{align}
In Rayleigh fading, $\tsnr\,z$ follows an exponential distribution with probability density function (pdf)
\begin{equation}
f(x)=\frac{1}{\tsnr\,\bar{z}}e^{-x/(\tsnr\,\bar{z})}.
\end{equation}
According to the results in \cite{akkouchi2008convolution}, the summation of independent exponentially distributed random variables $S_M=\sum_{k=1}^M X_k$, where $X_k\sim \exp(\lambda_k)$, has a pdf given by
\begin{align}
f_{S_M}(s)=\sum_{i=1}^{M}\frac{\prod_{j=1}^M\lambda_j}{\prod_{j=1,j\neq i}^M(\lambda_j-\lambda_i)}e^{-s\lambda_i}.
\end{align}
Using this characterization, the sum terms $\sum_{k\in\pi_i}\tsnr_k z_{kj}$ and $\sum_{k\in\pi_i, k\neq j}\tsnr_k z_{kj}$ in (\ref{eq:int_simplify}) can be regarded as two random variables, and the average rate can be evaluated using two single integrals. Similar approach can be applied to some other fading models as well.
\end{Rem}

~\\
\textbf{Mode Selection}

If a group just contains a single D2D direct link, then this D2D pair can choose between D2D mode and cellular mode. In cellular mode, D2D users communicate through the base station, and each time block is divided into two phases. In the first phase, the D2D transmitter sends packets to the base station, and base station forwards the packets to the corresponding D2D receiver in the second phase. We assume that the base station decodes and stores the received packets from D2D transmitters in a buffer, and the buffer empty probability is negligible. Let $\tau_i$ denote the fraction of time allocated to link $\bDT_i-\BS$. If users $\bDT_i$ and $\bDR_i$ are in cellular mode, then the fraction of time allocated to $\BS-\bDR_i$ link is $1-\tau_i$. Since the throughput of the two-hop link $\bDT_i-\BS-\bDR_i$ is $\min\{\tau_i\E\{R_{\bDT_i-\BS}\},(1-\tau_i)\E\{R_{\BS-\bDR_i}\}\}$, the optimal $\tau_i$ value is given by
\begin{align}\label{eq:opt_tau}
\tau_i^*=\frac{\E\{R_{\BS-\bDR_i}\}}{\E\{R_{\bDT_i-\BS}\}+\E\{R_{\BS-\bDR_i}\}},
\end{align}
which leads to $\tau_i\E\{R_{\bDT_i-\BS}\}=(1-\tau_i)\E\{R_{\BS-\bDR_i}\}$. Above in (\ref{eq:opt_tau}), the instantaneous rates of links $\bDT_i-\BS$ and $\BS-\bDR_i$ are formulated as
\begin{align}
R_{\bDT_i-\BS}&=B\log_2\left(1+\frac{P_d}{B\sigma^2}z_{\bDT_i-\BS}\right)\\
R_{\BS-\bDR_i}&=B\log_2\left(1+\frac{P_b}{B\sigma^2}z_{\BS-\bDR_i}\right)
\end{align}
where the subscript of the fading power $z$ denotes the link to which the fading power is associated. Then, the average transmission rate of the $i^{\text{th}}$ D2D pair in cellular mode is
\begin{align}
\E\{R_{\bDT_i-\BS-\bDR_i}\}&=\tau_i^*\E\{R_{\bDT_i-\BS}\}\\
&=\frac{\E\{R_{\BS-\bDR_i}\}\E\{R_{\bDT_i-\BS}\}}{\E\{R_{\bDT_i-\BS}\}+\E\{R_{\BS-\bDR_i}\}}.
\end{align}
In D2D mode, the average transmission rate of link $\bDT_i-\bDR_i$ is
\begin{align}
\E\{R_{\bDT_i-\bDR_i}\}=\E\left\{B\log_2\left(1+\frac{P_d}{B\sigma^2}z_{jj}\right)\right\}
\end{align}
where $j=2N_c+i$ is the index of the $i^{\text{th}}$ D2D direct link. We compare the average rates in these two modes, and select the one with the higher average rate.

\subsection{Channel Assignment}
In the first step, we divide the transmission links into $n_g$ groups, and the optimal transmission power and transmission mode of each user are obtained in the second step. In this third step discussed in this subsection, we allocate channel resources to each group.

We first allocate a channel to each group containing a cellular link, to guarantee that each cellular link is provided a channel. Following this step, there are $N-2N_c$ channels left for the remaining D2D users. Given these channels, we can choose to maximize the sum rate or maximize the total number of users served by the system.

If we choose to maximize the sum rate, then we need to pick $N-2N_c$ groups with the highest group sum rates from the remaining $n_g-2N_c$ groups, and assign each of them a channel. If we choose to maximize the number of users served by the system, then we need to select $N-2N_c$ groups with the largest group sizes, and assign each of them a channel.

\subsection{Summary}
Our joint mode selection and resource allocation algorithm is described in Table \ref{Alg:Overall}. Via the vertex coloring algorithm, we can quickly divide users into small groups, which greatly lowers the dimensionality of the power optimization problem in the second step and reduces the time consumption. From numerical results, we notice that the majority of the time is spent on solving the power optimization problems in the second step. Therefore, finding a faster algorithm instead of the interior-point method for the power optimization problem is the key to further reduce the time consumption of our algorithm, and we leave a detailed study of this problem as our future work.
\begin{table}
\footnotesize
\caption{\label{Alg:Overall}Algorithm 3}
\begin{tabular}{p{8.4cm}}
\hline
\hline
Joint Mode Selection and Resource Allocation Algorithm\\
\hline
\hline
Run Algorithm 2 for a given $\gamma$ value to obtain a partition with size $N_g$ greater or equal to the number of channels $N$;\\
\textbf{For} $i=1:n_g$\\
\hspace{0.5cm} Run the power optimization algorithm for the $i^{\text{th}}$ group;\\
\hspace{0.5cm} \textbf{If} the $i^{\text{th}}$ group only contains one D2D link\\
\hspace{1cm} Run the mode selection algorithm for this D2D link;\\
\hspace{0.5cm}\textbf{End}\\
\textbf{End}\\
Run the channel assignment algorithm to assign channel resources to these groups.\\
\hline
\hline
\end{tabular}
\end{table}

\section{Numerical Results}
\begin{figure*}
\begin{minipage}[b]{0.35\linewidth}
\begin{center}
\includegraphics[width=\textwidth]{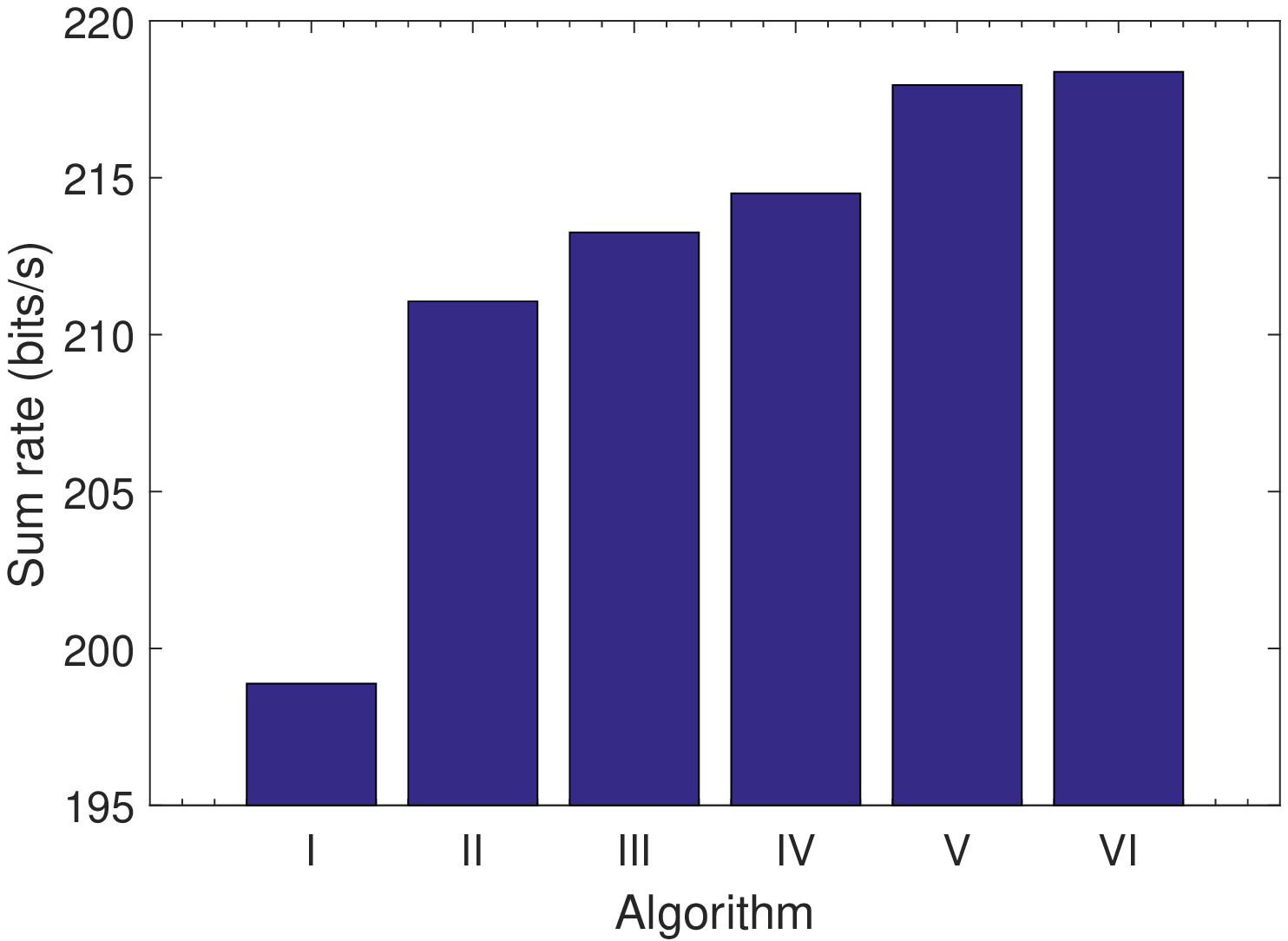}
\caption{Comparison of sum rate}\label{fig1}
\end{center}
\end{minipage}
\begin{minipage}[b]{0.35\linewidth}
\begin{center}
\includegraphics[width=\textwidth]{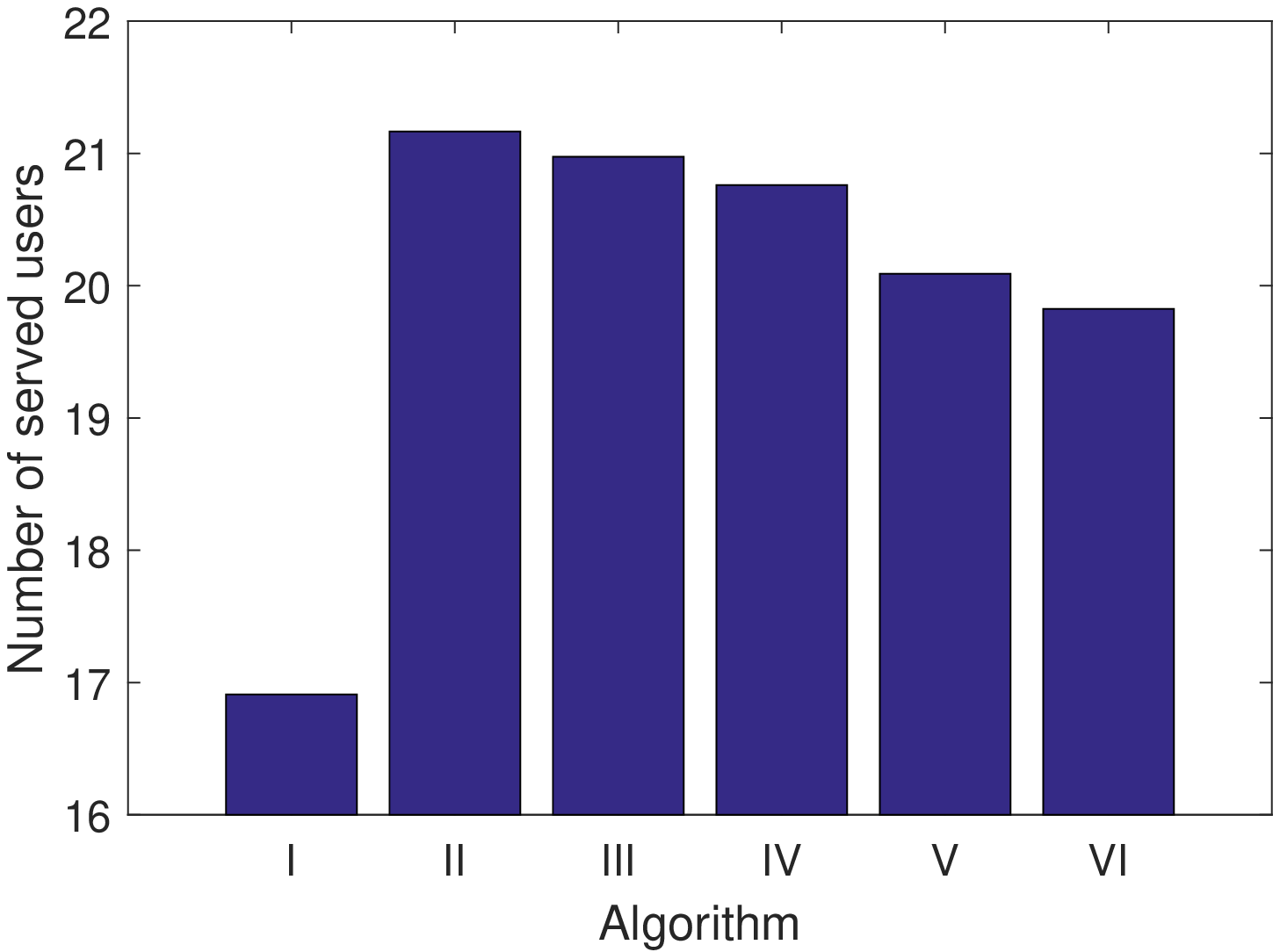}
\caption{Comparison of the number of served users}\label{fig2}
\end{center}
\end{minipage}
\begin{minipage}[b]{0.35\linewidth}
\begin{center}
\includegraphics[width=\textwidth]{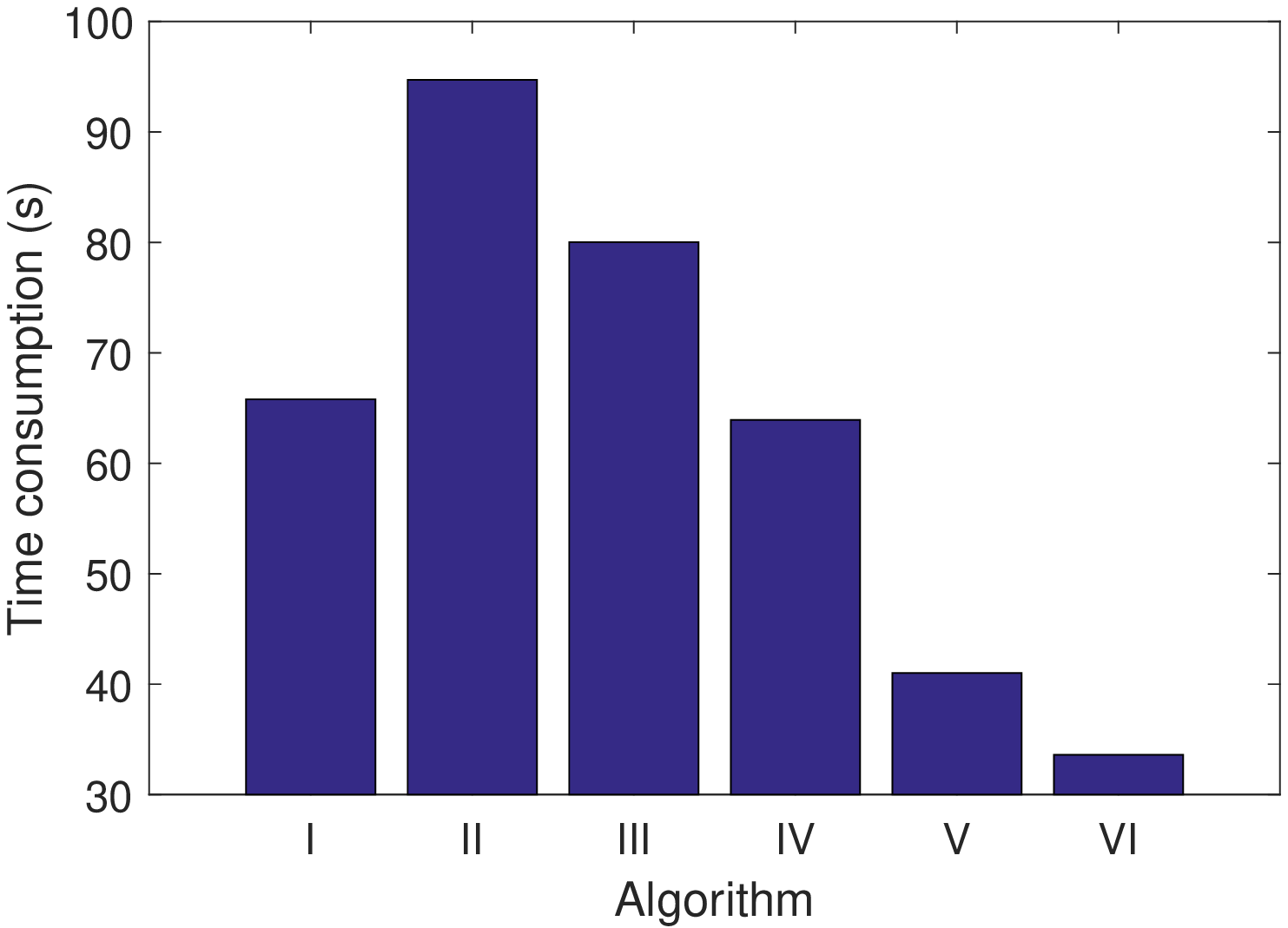}
\caption{Comparison of time consumption}\label{fig3}
\end{center}
\end{minipage}
\begin{minipage}[b]{0.5\linewidth}
\begin{center}
\includegraphics[width=0.9\textwidth]{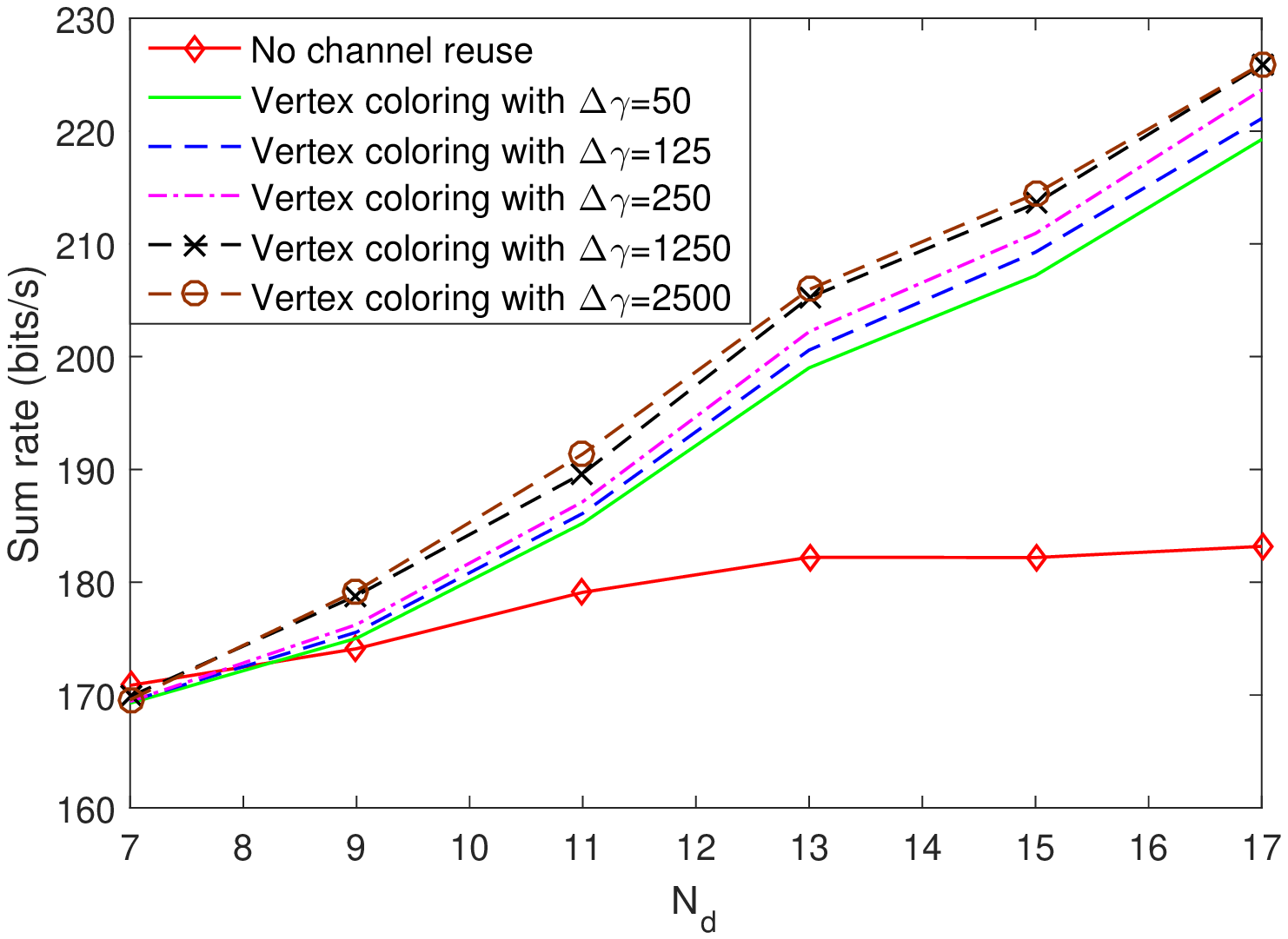}
\caption{Sum rate vs. $N_d$}\label{fig4}
\end{center}
\end{minipage}
\begin{minipage}[b]{0.5\linewidth}
\begin{center}
\includegraphics[width=0.9\textwidth]{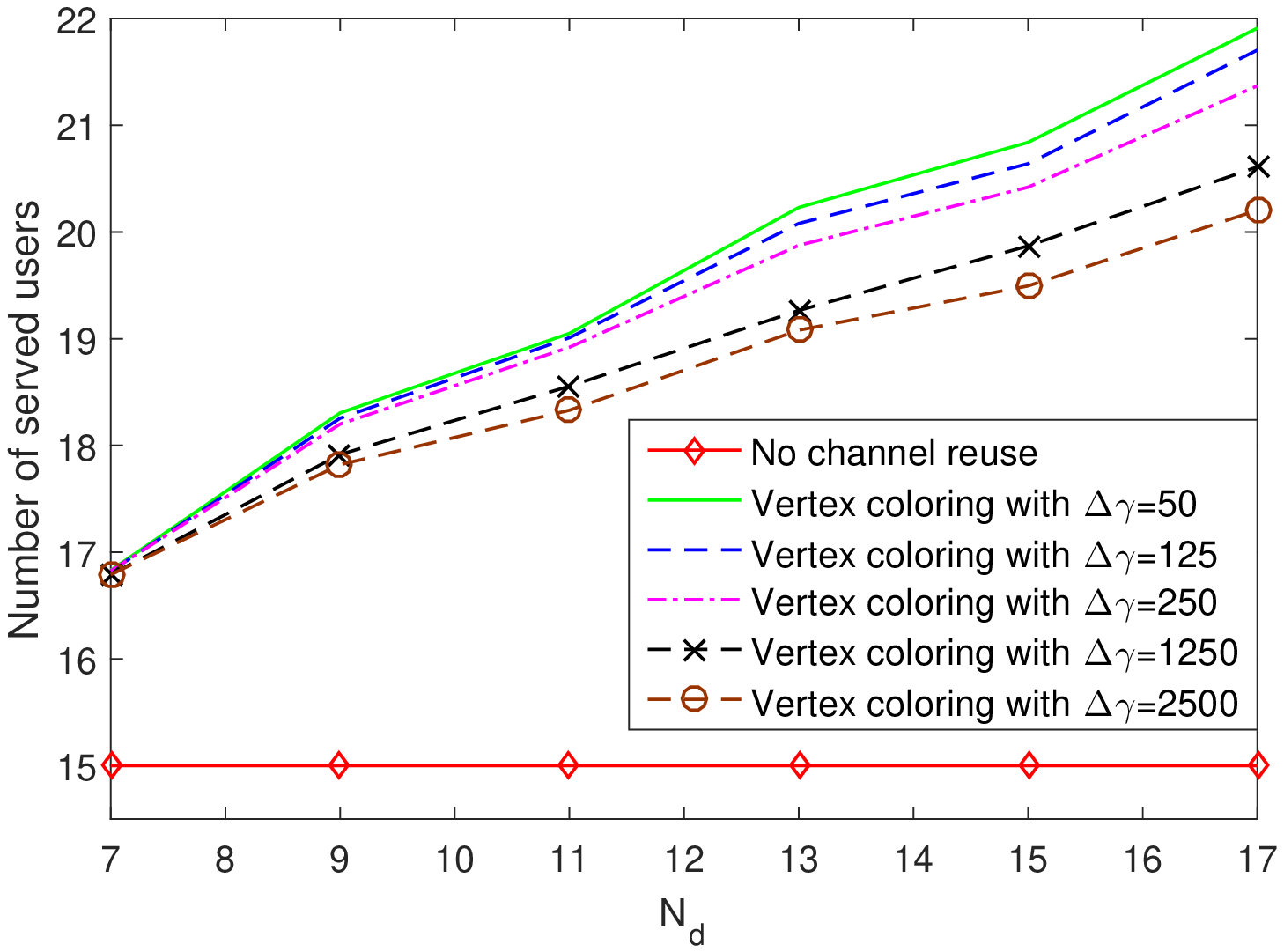}
\caption{Number of users served by the system vs. $N_d$}\label{fig5}
\end{center}
\end{minipage}
\end{figure*}
In this section, we further investigate the performance and parameter selection of our joint mode selection and resource allocation algorithm via simulations. For our algorithm, we set the initial threshold of the interference constraints as $\gamma=250$, and $\Delta\gamma\in\{50,125,250,1250,2500\}$. In the power allocation step, we set the weight for the minimum rate objective as $\mu=0.2\times\text{Size}(\pi_i)$, where size$(\pi_i)$ represents the number of links in the $i^{\text{th}}$ group. In the channel assignment step, we choose to maximize the sum rate. The number of channels is $N=25$, and the number of cellular users is $N_c=10$. The maximum SNRs are $\frac{P_b}{B\sigma^2}=27.78$ dB, $\frac{P_c}{B\sigma^2}=\frac{P_d}{B\sigma^2}=26.99$ dB. We consider Rayleigh fading with path loss $\E\{z\}=d^{-4}$, where $d$ is the transmission distance, and users are randomly placed in the cell. We repeat each simulation $200$ times, and each point in the numerical plots is averaged over $200$ randomly generated systems.

In Figs. \ref{fig1}-\ref{fig3}, we compare the performance of our algorithm with the coalitional game method proposed in \cite{li2014coalitional}. In the coalitional game, each user forms a coalition at the beginning, and link $i$ prefers to join coalition $j$ if the sum objective function increases by moving link $i$ to coalition $j$. In Figs. \ref{fig1}-\ref{fig3}, algorithms I, II, III, IV, V and VI represent the coalitional game algorithm and the vertex coloring algorithms with $\Delta\gamma=50$, $\Delta\gamma=125$, $\Delta\gamma=250$, $\Delta\gamma=1250$ and $\Delta\gamma=2500$, respectively, and the number of D2D pairs is fixed as $N_d=15$. From the results, we can see that our vertex coloring algorithm provides higher sum rates and serves more users than the coalitional game algorithm. As $\Delta\gamma$ increases, the sum rate increases due to less interference, but the number of users being served decreases due to stricter interference constraints, as noted in Section \ref{sec:partition}. Also, we notice that as $\Delta\gamma$ increases, the time consumption of our algorithm reduces, and when $\Delta\gamma=2500$, our algorithm is much faster than the coalitional game algorithm\footnote{These time consumption measurements are obtained for codes in Matlab 2015b running on a 2.40GHz Intel i7-4700MQ CPU.}. For larger values of $\Delta\gamma$, the maximum group size is small, reducing the dimensionality and time consumption of the power allocation problems in the second step.

In Figs. \ref{fig4} and \ref{fig5}, we plot the sum rate and number of served users as functions of the number of D2D pairs $N_d$. In these two figures, we consider the results without channel reuse as the benchmark, in which all users transmit with their maximum power and only $25$ links ($10$ cellular uplinks, $10$ cellular downlinks and $5$ D2D links) with the highest rates are allocated dedicated channels. We can see that as $N_d$ increases, the advantages of allowing channel reuse become more obvious. The sum rate and number of served users increase much faster when channel reuse is allowed. Similarly, larger $\Delta\gamma$ improves the sum rate while sacrificing the number of served users.

In summary, our algorithm has high performance and low time consumption. When the sum rate is more important, we can choose relatively high values for $\gamma$ and $\Delta\gamma$, and assign channels to the groups with higher group rates. In this case, the time consumption can also be reduced. However, if the values of $\gamma$ and $\Delta\gamma$ are too large, then our algorithm leads to the cases in which channel reuse is not allowed. On the other hand, we can choose relatively low values for $\gamma$ and $\Delta\gamma$, and assign channels to the groups with more users if we choose to maximize the number of served users. However, if the values of $\gamma$ and $\Delta\gamma$ are too small, then the interference limits the transmission rate of each user, and the service quality may degrade. Therefore, avoiding such extreme values and optimizing parameter selection are preferred.
\section{Conclusion}
In this work, we have proposed a joint mode selection and resource allocation algorithm for D2D underlaid cellular networks. We have decomposed the problem into three subproblems, and designed algorithms for each subproblem. In the first step, we divide the transmission links into small groups using vertex coloring algorithm. In the second step, we solve the power optimization problem using the interior-point method for each group and conduct mode selection for those D2D links which form a group, and we assign channel resources in the final step. Via simulation results, we have compared the performance of our algorithm with that of the coalitional game method, and have shown that our algorithm achieves higher sum rate and serves more users with relatively small time consumption. Also, the influence of the interference threshold step size $\Delta\gamma$ is studied through numerical results, and the tradeoff between sum rate and the number of served users is identified.

\bibliographystyle{ieeetr}
\bibliography{mode_selection}

\end{document}